\documentclass[12pt,preprint]{emulateapj}
\usepackage{subfigure}
\usepackage{natbib}
\usepackage{url}
\usepackage[backref,breaklinks,colorlinks,citecolor=blue]{hyperref}        
\usepackage[all]{hypcap}

\usepackage{color}
\begin{document}

\shorttitle{HiGAL - Spitzer Lifetimes}
\shortauthors{Battersby et al.}

\def\Msun{\hbox{M$_{\odot}$}}
\def\Lsun{\hbox{L$_{\odot}$}}
\def\kms{km~s$^{\rm -1}$}
\def\hcop{HCO$^{+}$}
\def\n2hp{N$_{2}$H$^{+}$}
\def\micron{$\mu$m}
\def\13CO{$^{13}$CO}
\def\etamb{$\eta_{\rm mb}$}
\def\Inu{I$_{\nu}$}
\def\kapnu{$\kappa _{\nu}$}
\def\ffore{f$_{\rm{fore}}$}
\def\tastar{T$_{A}^{*}$}
\def\nh3{NH$_{3}$}
\def\deg{$^{\circ}$}
\def\arcsec{$^{\prime\prime}$}
\def\arcmin{$^{\prime}$}
\def\Vlsr{\hbox{V$_{LSR}$}}

\title{The Lifetimes of Phases in High-Mass Star-Forming Regions}

\author{Cara Battersby\altaffilmark{1,2}, 
John Bally\altaffilmark{3}, \&
Brian Svoboda\altaffilmark{4}}

\altaffiltext{1}{Harvard-Smithsonian Center for Astrophysics, 60 Garden St., Cambridge, MA 02138, USA}
\altaffiltext{2}{University of Connecticut, Department of Physics, 2152 Hillside Road, Unit 3046
Storrs, CT 06269, USA}
\altaffiltext{3}{CASA, University of Colorado, 389 UCB, Boulder, CO 80309, USA}
\altaffiltext{4}{Steward Observatory, University of Arizona, 933 North Cherry Avenue, Tucson, AZ 85721, USA}

\begin{abstract}

High-mass stars form within star clusters from dense, molecular regions, but is the process of cluster formation slow and hydrostatic or quick and dynamic?  We link the physical properties of high-mass star-forming regions with their evolutionary stage in a systematic way, using Herschel and Spitzer data.  In order to produce a robust estimate of the relative lifetimes of these regions, we compare the fraction of dense, molecular regions above a column density associated with high-mass star formation, N(H$_2$) $>$ 0.4-2.5 $\times$ 10$^{22}$ cm$^{-2}$, in the `starless' (no signature of stars $\gtrsim$ 10 \Msun~forming) and star-forming phases in a 2\deg$\times$2\deg~region of the Galactic Plane centered at $\ell$=30\deg.  Of regions capable of forming high-mass stars on $\sim$1 pc scales, the starless (or embedded beyond detection) phase occupies about 60-70\% of the dense molecular region lifetime and the star-forming phase occupies about 30-40\%.  These relative lifetimes are robust over a wide range of thresholds.  We outline a method by which relative lifetimes can be anchored to absolute lifetimes from large-scale surveys of methanol masers and UCHII regions.  A simplistic application of this method estimates the absolute lifetime of the starless phase to be 0.2-1.7 Myr (about 0.6-4.1 fiducial cloud free-fall times) and the star-forming phase to be 0.1-0.7 Myr (about 0.4-2.4 free-fall times), but these are highly uncertain.  This work uniquely investigates the star-forming nature of high-column density gas pixel-by-pixel and our results demonstrate that the majority of high-column density gas is in a starless or embedded phase.

\end{abstract}

\keywords{ISM: kinematics and dynamics --- dust, extinction --- HII regions --- stars: formation}


\section{Introduction}
Whether star clusters and high-mass stars form as the result of slow, equilibrium collapse of clumps \citep[e.g.,][]{tan06} over several free-fall times or if they collapse quickly on the order of a free-fall time \citep[e.g.,][]{elm07, har07}, perhaps mediated by large scale accretion along filaments \citep{mye09}, remains an open question.
The stars that form in these regions may disrupt and re-distribute the molecular material from which they formed without dissociating it, allowing future generations of star formation in the cloud with overall long GMC lifetimes \citep[20-40 Myr, e.g.;][]{kaw09}.  
The scenario of quick, dynamic star formation sustained over a long time by continued inflow of material is motivated by a variety of observations \citep[discussed in detail in][]{elm07, har07}, and more recently by the lack of starless massive protoclusters \citep{gin12, urq14, cse14} observed through blind surveys of cold dust continuum emission in the Galaxy.  Additionally, observations of infall of molecular material on large scales \citep{sch10, per13} suggest that GMCs are dynamic and evolve quickly, but that material may be continually supplied into the region.  

To study the formation, early evolution, and lifetimes of high-mass star-forming regions, we investigate their earliest evolutionary phase in dense, molecular regions (DMRs).  The gas in regions that form high-mass stars, DMRs, has high densities \citep[10$^{4-7}$ cm$^{-3}$; ][]{lad03} and cold temperatures \citep[10-20 K; ][]{rat10, bat11} and is typically detected by submm observations where the cold, dust continuum emission peaks.
Given the appropriate viewing angle, these regions can also be seen in silhouette as Infrared Dark Clouds (IRDCs) absorbing the diffuse, mid-IR, Galactic background light.
\citet{bat11} showed that by using a combination of data \citep[including measurements of their temperatures and column densities from Hi-GAL;][]{mol10}, we can sample DMRs and classify them as starless or star-forming in a more systematically robust way than just using one wavelength (e.g. an IRDC would not be `dark' on the far side of the Galaxy, so a mid-IR-only selection would exclude those DMRs).  

\begin{figure*}
\centering
\includegraphics[trim=0mm 0mm 0mm 5mm, scale=0.75]{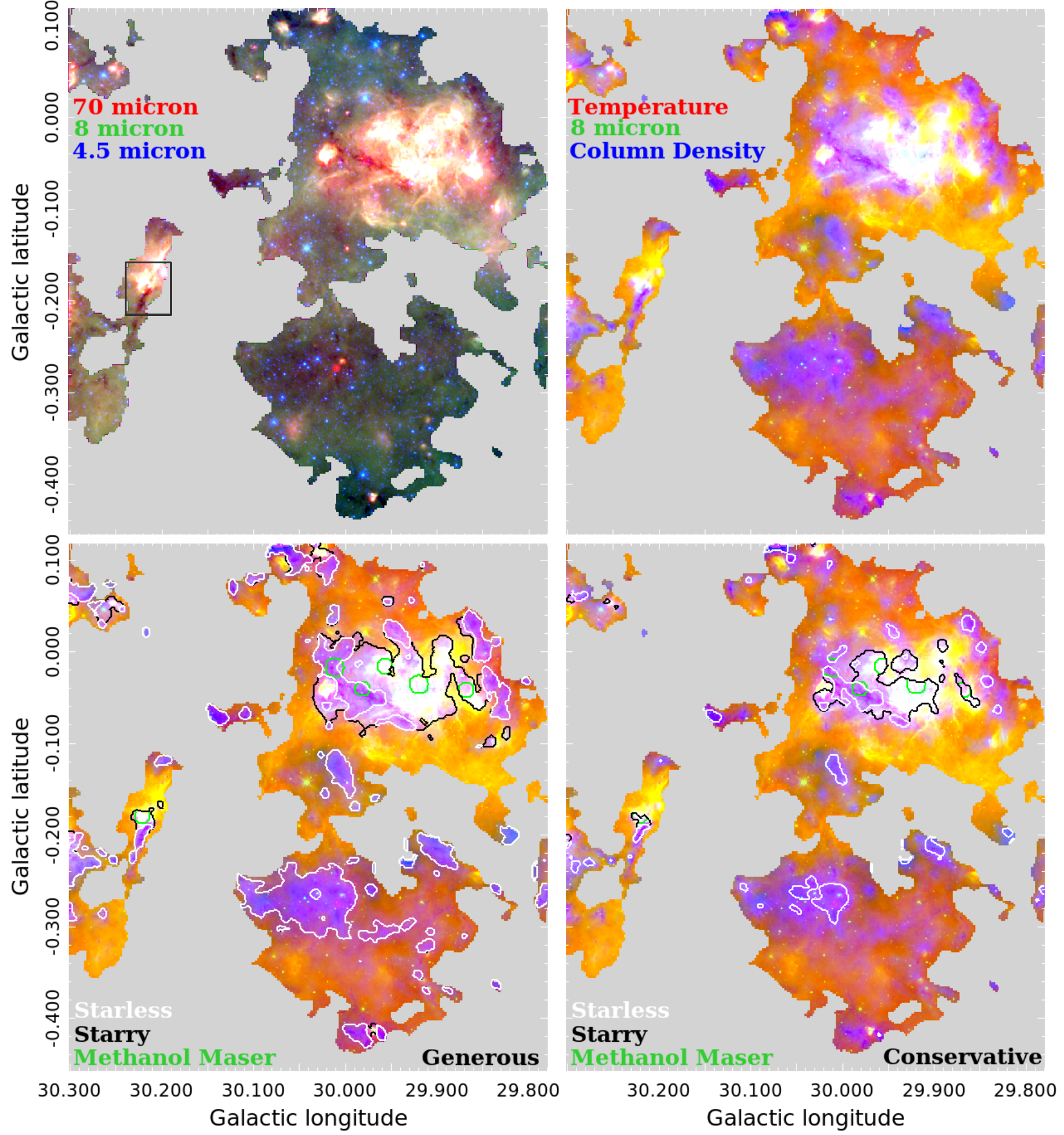}
\caption{A visual demonstration of the identification of DMRs and their classification as starless (pixels within the white contours) or star-forming (pixels within the black contours).  Shown in the \textit{top left} panel is a zoom-in map of a portion of our total field with red/green/blue = 70/8/4.5 \micron~emission.  Shown in the \textit{top right} panel and background throughout is the same zoom-in map with 
temperature shown in red, 8 \micron~emission in green, and column density in blue.
Predominantly blue regions in this map are starless DMRs (cold, high column density, and 8 \micron~dark), while predominantly white regions are star-forming DMRs (warm, high column density, and 8\micron~bright).  Red/yellow regions are warm but have low column densities and thus are not classified as DMRs.  In the \textit{bottom} panels, DMR identifications as starless or star-forming are shown in white or black contours, respectively, with the `generous' criteria on the left and `conservative' on the right.  The maser locations are shown by green contours.
Both identification methods yield relative lifetimes of approximately 70\% starless and 30\% star-forming, though the associated relative maser lifetime is 2\% on the left and 4\% on the right.  \citep[The sharp edges are source masks from][]{bat11}.
}
\label{fig:map}
\end{figure*}

Most previous studies of high-mass star-forming region lifetimes have focused on discrete `clumps' of gas, usually identified in the dust continuum \citep[e.g.][]{hey16,cse14, dun11a}.  However, oftentimes these `clumps' contain sub-regions of quiescence and active star formation and cannot simply be classified as either.  To lump these regions together and assign the entire `clump' as star-forming or quiescent can cause information to be lost on smaller scales within the clump.  For example, the filamentary cloud highlighted in a black box in \autoref{fig:map} centered at [30.21\deg, -0.18\deg] clearly contains an actively star-forming and a quiescent region, but is identified as a single clump in the Bolocam Galactic Plane Survey \citep{gin13, ros10, agu11}, 
as 3 clumps in the first ATLASGAL catalog \citep{contreras13, urq14b}, and as 5 clumps in the second ATLASGAL catalog \citep{cse14}.  Resolution and sensitivity are not the primary drivers for the number of clumps identified, rather it is algorithmic differences, such as bolocat vs. clumpfind or gaussclumps. 

In this paper, we present an alternate approach that circumvents the issues surrounding clump identification, by retaining the available information in a pixel-by-pixel analysis of the maps.  All together, the pixels give us a statistical overview of the different evolutionary stages.  Pixels that satisfy the criteria for high-mass star formation, explicated in the paper, are referred to as dense molecular regions (DMRs).  We compare the fractions of the statistical ensembles of starless and star-forming DMRs to estimate their relative lifetimes.

Previous lifetime estimates, based primarily on mid-IR emission signatures toward samples of IRDCs or dust continuum clumps, found relative starless fractions between about 30-80\% and extrapolate these to absolute starless lifetimes ranging anywhere between 10$^{3}$-10$^{6}$ years.  
Notably, the recent comprehensive studies of clump lifetimes from \citet{svo16} and \citet{hey16} find starless fractions of 47\% and 69\%, respectively.
Our approach differs from previous methods by 1) not introducing clump boundaries, thereby using the highest resolution information available, 2) defining regions capable of forming high-mass stars based strictly on their column density, and 3) using the dust temperature and the mid-IR emission signature to classify a region as starless or star-forming.  

\section{Methods}
\label{sec:method}

The Galactic Plane centered at Galactic longitude $\ell =$ 30\arcdeg\ contains one of the largest concentrations of dense gas and dust in the Milky Way.  Located near the end of the Galactic bar and the start of the Scutum-Centaurus spiral arm, this region contains the massive W43 star forming complex at a distance of 5.5 kpc \citep{ell15} and hundreds of massive clumps and molecular clouds with more than 80\% of the emission having \Vlsr\ =  80 to 120 \kms\ implying a kinematic distance between 5 and 9 kpc \citep{Carlhoff13, ell13,ell15}.    
We investigate the properties and star-forming stages of DMRs in a 2\deg $\times$ 2\deg~field centered at [$\ell$,b] = [30\deg, 0\deg] using survey data from Hi-GAL \citep{mol10}, GLIMPSE \citep{ben03}, 6.7 GHz CH$_{3}$OH masers \citep{pes05}, and UCHII regions \citep{woo89}.

In a previous work, we measured T$_{dust}$ and N(H$_{2}$) from modified blackbody fits to background-subtracted Hi-GAL data from 160 to 500 \micron~using methods described in \citet{bat11} and identified each pixel as mid-IR-bright, mid-IR-neutral, or mid-IR-dark, based on its contrast at 8 \micron.  The column density and temperature maps are produced using data which have had the diffuse Galactic component removed, as described in \citet{bat11}.  We use the 25\arcsec~version of the maps, convolving and regridding all the data to this resolution, which corresponds to beam sizes of 0.6 and 1.1pc and pixel sizes of 0.13 and 0.24 pc at typical distances of 5 and 9 kpc.
  
In this work, we identify pixels above a column density threshold capable of forming high-mass stars (\S \ref{sec:nh2_thresh}), then, on a pixel-by-pixel basis, identify them as starless or star-forming (\S \ref{sec:starry}), and from their fractions infer their relative lifetimes.  Using absolute lifetimes estimated from survey statistics of 6.7 CH$_{3}$OH masers and UCHII regions, we anchor our relative lifetimes to estimate the absolute lifetimes of the DMRs (\S \ref{sec:maser} and \ref{sec:uchii}).

Previous works (see \S \ref{sec:comp}) have estimated star-forming lifetimes over contiguous ``clumps," typically $\sim$1 pc in size.  However, sub-mm identified clumps often contain distinct regions in different stages of star formation.  It was this realization that led us to the pixel-by-pixel approach.  In the clump approach, actively star-forming and quiescent gas are lumped together; a single signature of star formation in a clump will qualify all of the gas within it as star-forming.  This association of a large amount of non-star-forming gas with a star-forming clump could lead to erroneous lifetime estimates for the phases of high-mass star formation, though we note that this will be less problematic in higher-resolution analyses.  Therefore, we use a pixel-by-pixel approach and consider any pixel with sufficient column density (see next section) to be a dense molecular region, DMR.

%
%

 \subsection{Column Density Thresholds for High-Mass Star Formation}
 \label{sec:nh2_thresh}
 
High-mass stars form in regions with surface densities $\Sigma$ $\sim$ 1 g cm$^{-2}$ \citep{kru08}, corresponding to N(H$_{2}$) $\sim$ 2.1$\times$10$^{23}$ cm$^{-2}$.    At distances of several kpc or more, most cores are highly beam-diluted in our 25\arcsec\ beam.  To derive a realistic high-mass star-forming column density threshold for cores beam-diluted by a 25\arcsec\ beam, consider a spherical core with a constant column density $\Sigma$ = 1 g cm$^{-2}$ in the central core (defined as $r <  r_{f}$, where r$_f$ is the flat inner portion of the density profile, determined from fits to the data) and a power-law drop off for  $r >  r_{f}$   
\begin{equation}
n(r) = n_{f} (r / r_{f}) ^{-p}
\end{equation}
where p is the density power-law exponent.  The \citet{mue02} study of 51 high-mass star-forming cores found a best-fit central core radius, r$_{f} ~\approx$ 1000 AU, and density power-law index, $p$ = 1.8.  This model implies an H$_2$ central density of  $n_{f}$ = 6.2 $\times$ 10$^{7}$ cm$^{-3}$, which, integrated over $r_{f}$ =  1000 AU, corresponds to the theoretical surface density threshold for forming high-mass stars of $\Sigma$ = 1 g cm$^{-2}$.

Integration of this model core along the line of sight and convolution with a 25\arcsec\ beam results in a beam-diluted column density threshold, at typical distances of 5 and 9 kpc toward the $\ell$ = 30\deg~field \citep{ell13, ell15}, of N(H$_{2}$) = 0.8 and 0.4 $\times$ 10$^{22}$ cm$^{-2}$, respectively.  Pixels above this column density threshold are referred to as dense molecular regions (DMRs).  We note that the column density maps used have had the diffuse Galactic background removed, as described in \citet{bat11}, so we can attribute all of the column density to the DMRs themselves.

As discussed in \S \ref{sec:colfig}, the relative lifetimes are mostly insensitive to variations in the threshold column density from $\sim$0.3 to 1.3 $\times$ 10$^{22}$ cm$^{-2}$ 
(corresponding to distances of 11 and 3 kpc for the model core).
We make two estimates of lifetimes throughout the text using the extreme ends of reasonable parameter space, for this section, the cutoffs are:   
N(H$_{2}$) = 0.4 $\times$ 10$^{22}$ cm$^{-2}$ for the `generous' and N(H$_{2}$) = 0.8 $\times$ 10$^{22}$ cm$^{-2}$ and `conservative' estimates.        

Alternatively, we apply the \citet{kau10} criteria (hereafter KP10) for high-mass star formation.  \citet{kau10} observationally find that regions with high-mass star formation tend to have a mass-radius relationship of m(r) $>$ 870 \Msun [r/pc]$^{1.33}$.  At our beam-sizes of 0.6 and 1.1 pc, this corresponds to column densities of about 2.5 and 1.6 $\times$ 10$^{22}$ cm$^{-2}$, respectively.  The results of this column density threshold are discussed in more detail in each results section (\S \ref{sec:colfig}, \ref{sec:maser}, and \ref{sec:uchii}).  The result that our relative lifetimes are mostly insensitive to variations in the the threshold column density holds true with these higher column density thresholds.

\begin{figure*}
\centering
\includegraphics[width=1\textwidth]{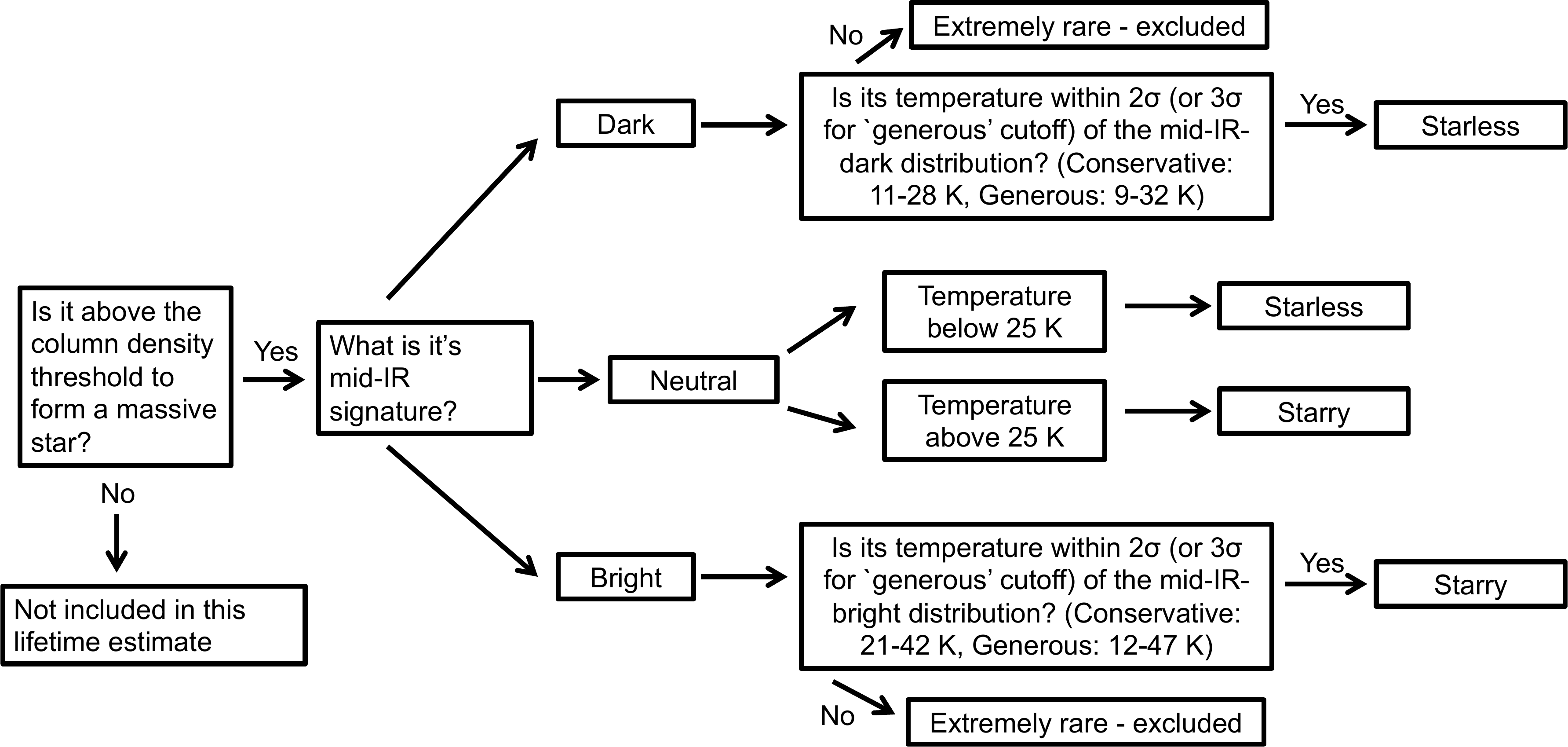}
\caption{Flow chart showing the decision tree to assign pixels as DMRs, and subsequently classify them as `starless' or `starry,' i.e. star-forming.
}
\label{fig:flow}
\end{figure*}

\subsection{Starless vs. Star-Forming}
\label{sec:starry}

In this paper, dust temperature distributions are combined with mid-IR star formation signatures to determine whether each DMR is `starless' or `star-forming'.  The mid-IR signature is determined from the contrast at 8 \micron, i.e. how `bright' or `dark' the pixel is above the background - the 8 \micron~image smoothed with a median filter of 25\arcsec\ resolution \citep[see][for details]{bat11}. 
We use the mid-IR signature at 8 \micron~(mid-IR dark or bright) as the main discriminator and the dust temperature as a secondary discriminator, particularly to help identify regions that are cold and starless, but do not show particular absorption as an IRDC as they be on the far side of the Galaxy.

We use the range of dust temperatures found to be associated with mid-IR-dark and bright regions (based on Gaussian fits to those temperature distributions) to help discriminate DMRs are `starless' or `star-forming.'  If a DMR is mid-IR-dark and its temperature is within the normal cold dust range (2 or 3-$\sigma$ for the `conservative' and `generous' thresholds respectively), then it is classified as `starless.'
If a DMR is mid-IR-bright and its temperature is within the normal warm dust range (2 or 3-$\sigma$ for the `conservative' and `generous' thresholds respectively), then it is classified as `star-forming.'  Slight changes in the temperature distributions (e.g., including all DMRs down to 0 K as starless and up to 100 K as star-forming) have a negligible effect.  Pixels that are mid-IR-bright and cold or mid-IR-dark and warm are extremely rare and left out of the remaining analysis.  When a DMR is mid-IR-neutral, its temperature is used to classify it as `starless' or `star-forming.' 
A flow chart depicting this decision tree is shown in \autoref{fig:flow}.

This study is only sensitive to the star-forming signatures of high-mass stars.  Therefore, the term `starless' refers only to the absence of high-mass stars forming; the region may support active low- or intermediate-mass star formation.  Using \citet{rob06} models of dust emission toward YSOs, scaled to 5 and 9 kpc distances and apertures, we find that our typical starless flux limit of 130 MJy/Sr at 8 \micron~(technically we use a contrast not a specific flux, but this is the approximate level for most of the region at our contrast cut) is sensitive enough to detect the vast majority (85\%) of possible model YSOs with a mass above 10 \Msun.  Some YSO models will always be undetected (due to unfortunate inclinations, etc.) no matter the flux limit, so we estimate that we are sensitive to forming massive stars above $\sim$10 \Msun.

While contrast at 8 \micron~is an imperfect measure of the presence or absence of high-mass star formation, it was found to be a powerful discriminator in \citet{bat11}, who compared it with dust temperature, 24 \micron~emission, maser emission, and Extended Green Objects.  In particular, since we employ the contrast at 8 \micron~smoothed over 25\arcsec\ using a median filter, rather than the simple presence or absence of emission, we are unlikely to be affected by field stars, which are small and very bright, and thus will be removed by the median filter smoothing.  PAH emission at 8 \micron~toward Photon-Dominated Regions (PDRs) is not likely to be an issue since the low column density of dust towards PDRs will preclude their inclusion as DMRs in the first place.

\section{Lifetimes}
\label{sec:lifetimes}

\subsection{Assumptions in Deriving Relative Lifetimes}
\label{sec:caveats}	

Several necessary assumptions were made in inferring the relative lifetimes of the starless and star-forming stages for high-mass star-forming regions.  i) The sample is complete and unbiased in time and space.  ii) The sum of DMRs represents with equal probability all the phases in the formation of a massive star.  This can be achieved either with a constant star formation rate (SFR) or a large enough sample. iii) DMRs are forming or will form high-mass stars.  iv) The lifetimes don't depend on mass.  v) Starless and star-forming regions occupy similar areas on the sky (i.e. their area is proportional to the mass of gas).  vi) The signatures at 8 \micron~(mid-IR bright or dark) and their associated temperature distributions are good indicators of the presence or absence of a high-mass star.  vii) Pixels above the threshold column density contain beam-diluted dense cores rather than purely beam-filling low-surface density gas.  We discuss these assumptions below.

Assumptions (i) and (ii) are reasonable given the size of our region; we are sampling many populations in different evolutionary stages.  The relative lifetimes we derive here may apply elsewhere to special regions in the Galaxy with similar SFRs and properties.  However, since this region contains W43 a `mini-starburst' \citep[e.g.][]{bal10, lou14}, it is likely in a higher SFR phase and may not be applicable generally throughout the Galaxy.   
One possible complication to assumption (ii) is the possibility that some high-mass stars can be ejected from their birthplaces with velocities sufficient to be physically separated from their natal clumps within about 1 Myr.  However, this should not result in an underestimate of the star-forming fraction of DMRs, since high-mass star formation is highly clustered, and if the stellar density is sufficient for ejection of high-mass stars, then the remaining stars should easily classify the region as a DMR.  On the other hand, the ejected stars could in principle classify a dense, starless region they encounter as `star-forming' by proxy.  We expect that this would be rare, but suggest it as an uncertainty worthy of further investigation.

We argue in favor of assumption (iii) in \S \ref{sec:nh2_thresh}.
While we expect that assumption (iv) is not valid over all size clumps, this assumption is reasonable (and necessary) for our sample, as the column density variation over our DMRs is very small (from the threshold density to double that value).  This lack of column density variation in our DMRs also argues in favor of assumption (v). 
Various studies \citep[e.g.,][]{bat10, cha09} argue in favor of assumption (vi), statistically, but more sensitive and higher resolution studies will continue to shed light on the validity of this assumption.  Assumption (vii) is supported by the fact that interferometric observations of DMRs \citep[e.g.][]{bat14b} and density measurements \citep{gin15} demonstrate that most (if not all) DMRs contain dense substructure rather than purely beam-filling low surface-density gas.  While we expect that the high column density of some DMRs may not indicate a high volume density, but rather long filaments seen `pole-on,' we expect this to be quite rare.  If this special geometry were common, it would lead to an over-estimate of the starless fraction.

\begin{figure*}
\centering
\includegraphics[scale=0.75]{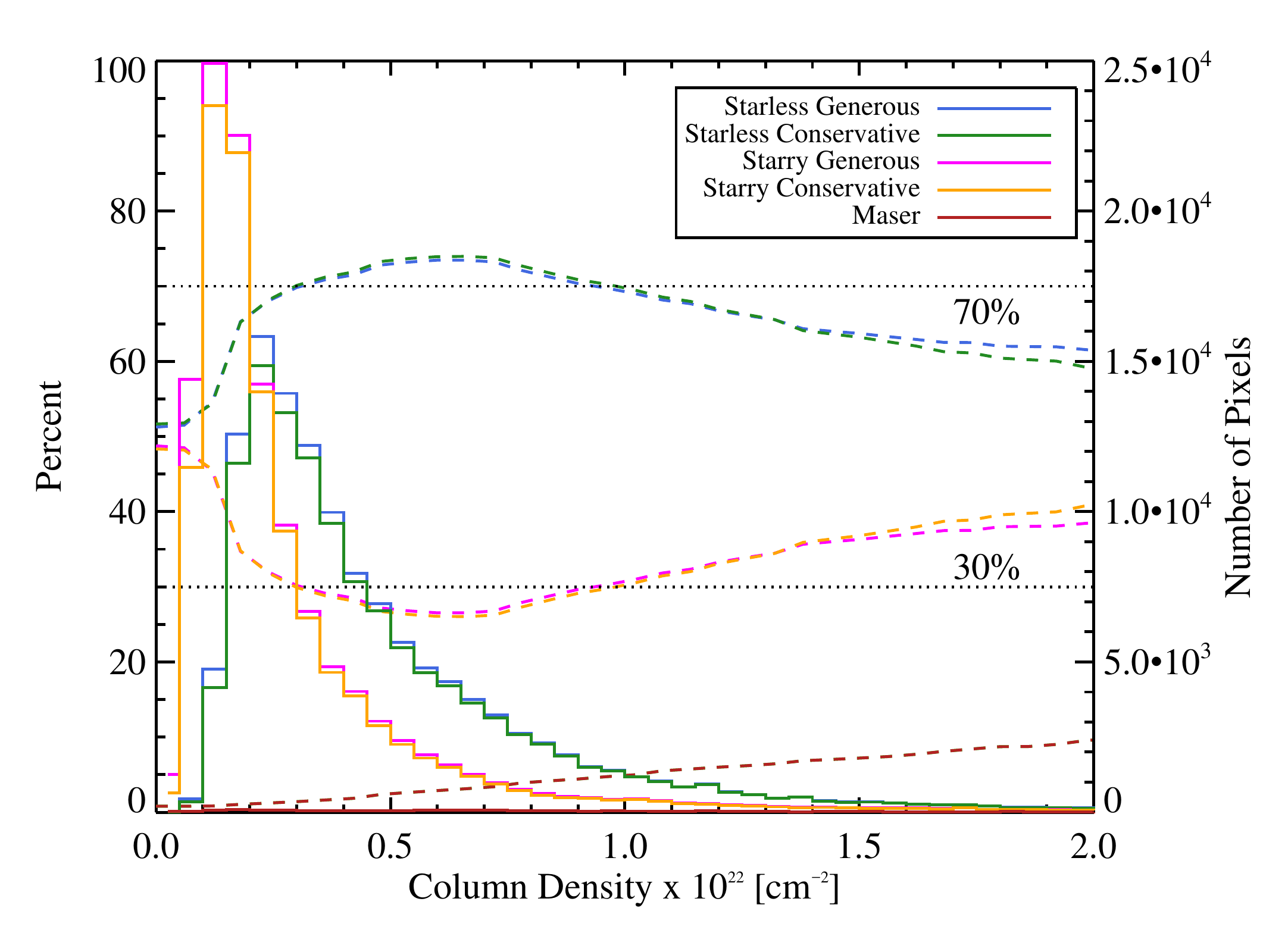}
\caption{
The relative fraction of pixels which are starless (70\%) vs. star-forming (30\%) is shown in the dashed lines as a function of column density cutoff.  This fraction is mostly insensitive to the choice of column density cutoff over a reasonable range of thresholds (N(H$_{2}$) $\sim$ 0.3 - 1.3 $\times$ 10$^{22}$ cm$^{-2}$) in a 25\arcsec\ beam and to the `generous' vs. `conservative' parameters.  The left y-axis and dashed lines show the percentage of pixels which are starless (cyan and green using the `generous' and `conservative' identification methods) vs. star-forming (magenta and orange using the `generous' and `conservative' identification methods) or which contain a maser (red) as a function of the column density threshold selected (x-axis).  The right y-axis and solid lines show the column density distribution of the same populations.  In each column density bin (x-axis), the ratio between the number of pixels in each category in that column density bin (solid lines) divided by the total number of pixels in that column density bin, multiplied by 100 gives the percentage of pixels in each category (dashed lines). This figure is discussed in more detail in \S \ref{sec:colfig}.
}
\label{fig:colcutoff}
\end{figure*}

\subsection{Observed Relative Lifetimes}
 \label{sec:colfig}

For both the `conservative' and `generous' estimates, the relative percentage of DMRs in the starless / star-forming phase is about 70\%/30\%.  The higher column density thresholds from KP10 give relative percentages of 63\%/37\%.  These lifetimes are shown in \autoref{table-comp}.
The dashed lines in \autoref{fig:colcutoff} shows the percentage of pixels in the starless vs. starry categories for a range of column density thresholds shown on the x-axis.  Below a column density threshold of about 0.3 $\times$ 10$^{22}$ cm$^{-2}$ starless and starry pixels are equally distributed (50\%).  At any column density threshold in the range of 0.3 - 1.3 $\times$ 10$^{22}$ cm$^{-2}$ about 70\% of the pixels are categorized as starless and 30\% as starry (i.e. star-forming).  At the higher column density thresholds from KP10, 1.7 - 2.5 $\times$ 10$^{22}$ cm$^{-2}$ about 63\% are starless and 37\% star-forming.  

At each column density, the ratio between the number of pixels in each category (solid lines in \autoref{fig:colcutoff}, N$_{starless}$ and N$_{starry}$) divided by the total number of pixels at that column density (N$_{total}$), multiplied by 100, gives the percentage of pixels in each category (dashed lines in \autoref{fig:colcutoff}, Starless~\% and Starry~\%).
\begin{equation}
\rm{Starless~\%} = [ N_{starless} / N_{total} ] \times 100 
\end{equation}
\begin{equation}
\rm{Starry~\%} = [ N_{starry} / N_{total} ] \times 100
\end{equation}

The histograms in solid lines in \autoref{fig:colcutoff} show the number of pixels in each category with a given column density, as noted on the x-axis; i.e. a column density probability distribution function.  This figure demonstrates that the average column density distribution is higher for starless pixels.  We interpret this to mean that regions categorized as starless have a high capacity for forming future stars (high column density and cold).
Our method of comparing these populations above a threshold column density allows us to disentangle them, and derive relative lifetime estimates for regions on the brink of (starless) vs. actively forming stars (starry).

Under the assumptions discussed in \S \ref{sec:method} and \ref{sec:caveats}, high-mass star-forming DMRs spend about 70\% of their lives in the starless phase and 30\% in the actively star-forming phase.
If we instead apply the KP10 criteria as a column density threshold (see \S \ref{sec:nh2_thresh}), our relative lifetimes are 63\% for the starless phase and 37\% the for star-forming phase.  We therefore conclude that the starless phase occupies approximately 60-70\% of the lifetime of DMRs while the star-forming phase is about 30-40\% over a wide range of parameters, both column density threshold and temperature thresholds as shown by `conservative' vs. `generous' criteria.
 
While our relative lifetime estimates are robust over a range of parameters, the connection of our relative lifetimes to absolute lifetimes is extremely uncertain.  We present below two methods to connect our relative lifetimes to absolute timescales.  The first is to link the methanol masers detected in our region with Galactic-scale maser surveys, which provide an estimate of the maser lifetime.  The second approach is to instead assume an absolute lifetime of UCHII regions to anchor our relative DMR lifetimes.

\subsection{Maser association and absolute lifetime estimates}
\label{sec:maser}

We use the association of DMRs with 6.7 GHz Class II methanol masers (thought to be almost exclusively associated with regions of high-mass star formation), and the lifetime of these masers from \citet{van05}, to anchor our relative lifetimes to absolute timescales.  These lifetimes are highly uncertain and rest on a number of assumptions, therefore care should be taken in their interpretation.  We utilize unbiased Galactic plane searches for methanol masers by \citet{szy02} and \citet{ell96} compiled by \citet{pes05}.  
We define the `sizes' of the methanol masers to be spherical regions with a radius determined by the average size of a cluster forming clump associated with a methanol maser from the BGPS as $R\sim0.5$ pc \citep{svo16}, corresponding to 30\arcsec~diameter apertures for the average distance in this field.
While methanol maser emission comes from very small areas of the sky \citep[e.g.,][]{wal98}, they are often clustered, so these methanol maser ``sizes" are meant to represent the extent of the star-forming region. 

The absolute lifetime of DMRs can be anchored to the duration of the 6.7 GHz Class II CH$_{3}$OH masers, which are estimated to have lifetimes of $\sim$35,000 years \citep[by extrapolating the number of masers identified in these same surveys to the total number of masers in the Milky Way and using an IMF and global SFR to estimate their lifetimes;][]{van05}.
We note that the 
\citet{van05} extrapolated total number of methanol masers in the Galaxy of 1200 is in surprisingly good agreement with the more recent published methanol maser count from the MMB group \citep[they find 582 over half the Galaxy;][]{cas10,cas11,gre12}, therefore, though this catalog is outdated, its absolute lifetime estimate remains intact, though we suggest that future works use these new MMB catalogs, which have exquisite positional accuracy.

While the fraction of starless vs. star-forming DMRs is insensitive to the column density cuts, the fraction of DMRs associated with methanol masers ($f_{maser}$) increases as a function of column density (see \autoref{fig:colcutoff}).  In the `generous' and `conservative' cuts, the methanol maser fraction is 2\% and 4\% ($f_{maser}$), respectively, corresponding to total DMR lifetimes ($\tau_{total}$) of 1.9 Myr and 0.9 Myr.  Using the alternative column density threshold from \citet{kau10}, called KP10, as discussed in \S \ref{sec:nh2_thresh}, the maser fraction is about 7-12\%, corresponding to total DMR lifetimes ($\tau_{total}$) of 0.3 and 0.4 Myr.  Given our starless and star-forming fractions ($f_{starless}$=0.6-0.7 and $f_{starry}$=0.3-0.4) and the methanol maser lifetime \citep[$\tau_{maser}$=35,000 years;][]{van05} we can calculate the total DMC lifetime and relative phase lifetimes using the following equations:
\begin{equation}
\tau_{total} = \frac{\tau_{maser}}{f_{maser}}
\end{equation}
\begin{equation}
\tau_{starless} = f_{starless}~ \tau_{total}
\end{equation}
\begin{equation}
\tau_{starry} = f_{starry}~ \tau_{total}
\end{equation}
The `starless' lifetime then is about 0.6-1.4 Myr, while the `star-forming' lifetime is 0.2-0.6 Myr considering only the `conservative' and `generous' thresholds.  The KP10 column density threshold for high-mass star formation, because of its larger methanol maser fraction, yields absolute starless lifetimes of 0.2-0.3 Myr and star-forming lifetimes of 0.1-0.2 Myr.  Overall, the range of absolute starless lifetimes is 0.2 - 1.4 Myr and star-forming lifetimes is 0.1 - 0.6 Myr.  See \autoref{table-comp} for a summary of the relative and absolute lifetimes for various methods.

\begin{table*}
\begin{center}
\begin{tabular}{llccccccc}
\hline
Criteria & N(H$_2$) [cm$^{-2}$] & $f_{starless}$ & $f_{starry}$ & Anchor & $\tau_{total}$ [Myr] & $\tau_{starless}$ & $\tau_{starry}$ \\ 
\hline\hline
Generous & 0.4$\times$10$^{22}$ & 0.71 & 0.29 & maser & 1.9  & 1.4  & 0.6  \\
Conservative & 0.8$\times$10$^{22}$ & 0.72 & 0.28 & maser & 0.9  & 0.6  & 0.3  \\
Either & 0.4-0.8$\times$10$^{22}$ & 0.7 & 0.3 & UCHII & 1.2-2.4  & 0.8-1.7  & 0.4-0.7  \\
KP10 near & 2.5$\times$10$^{22}$ & 0.63 & 0.37 & maser & 0.3  & 0.2  & 0.1  \\
KP10 far & 1.6$\times$10$^{22}$ & 0.63 & 0.37 & maser & 0.5  & 0.3  & 0.2  \\
Either KP & 1.6-2.5$\times$10$^{22}$ & 0.63 & 0.37 & UCHII & 1.0-1.9  & 0.6-1.2  & 0.4-0.7  \\
\hline
Overall & 0.4-2.5$\times$10$^{22}$ & 0.6-0.7 & 0.3-0.4 & both & 0.3-2.4 & 0.2-1.7  & 0.1-0.7  \\
\hline
\end{tabular}
\end{center}
\caption{Various lifetime estimates in this work, described in more detail in \S \ref{sec:method}, \ref{sec:colfig}, \ref{sec:maser}, and \ref{sec:uchii}.  The column density listed refers to the lower bound on the column densities used to derive the relative lifetimes.  The Anchor is the method used to anchor relative lifetimes to absolute timescales, CH$_3$OH masers or UCHII Regions.  Finally, the timescales in the three right-most columns are lifetimes of various phases in Myr.}
\label{table-comp}
\end{table*}

\begin{table*}
\begin{center}
\begin{tabular}{lllllll}
\hline
Criteria &  Median N(H$_2$) & Anchor & $\tau_{ff}$ [Myr] & N$_{ff, total}$ & N$_{ff, starless}$ &N$_{ff, starry}$ \\ 
\hline\hline
Generous &  0.6$\times$10$^{22}$ & maser & 0.7 & 2.8 & 2.0 & 0.8 \\
Conservative & 1.2$\times$10$^{22}$ & maser & 0.5  &1.8  & 1.3 & 0.5 \\
Either &  0.6-1.2$\times$10$^{22}$ & UCHII & 0.5-0.7  & 1.7-4.8 & 1.2-3.4 & 0.5-1.4  \\
KP10 near & 3.5$\times$10$^{22}$ & maser & 0.3  & 1.0 & 0.6  & 0.4 \\
KP10 far & 2.2$\times$10$^{22}$ & maser & 0.4 & 1.1 & 0.7 & 0.4  \\
Either KP &  2.2-3.5$\times$10$^{22}$& UCHII & 0.3-0.4 & 2.4-6.5 & 1.5-4.1 & 0.9-2.4 \\
\hline
Overall &  0.6-3.5$\times$10$^{22}$& both & 0.3-0.7 & 1.0-6.5 & 0.6-4.1 & 0.4-2.4 \\
\hline
\end{tabular}
\end{center}
\caption{Various lifetime estimates in this work, as in \autoref{table-comp}, in units of the number of free-fall times.  A robust estimate of the free-fall time is not performed in this work, these are simply based on simple geometrical assumptions and the median column density of pixels included in each estimate.  These are discussed in more detail in \S \ref{sec:ff}.}
\label{table-ff}
\end{table*}

\subsection{UCHII Region Association and Lifetimes}
\label{sec:uchii}

Since the absolute lifetimes of methanol masers is quite uncertain, we tie our relative lifetimes to absolute lifetimes of UCHII regions using a different method to probe the range of parameter space that is likely for DMRs.
\citet{woo89b,woo89} determined that the lifetimes of UCHII regions are longer than anticipated based on the expected expansion rate of D-type ionization fronts as HII regions evolve toward pressure equilibrium.  They estimate that O stars spend about 10-20\% of their main-sequence lifetime in molecular clouds as UCHII regions, or about 3.6 $\times$ 10$^{5}$ years\footnote{For an O6 star, the main sequence lifetime is about 2.4 $\times$ 10$^{6}$ years \citep{mae87}, so 15\% is 3.6 $\times$ 10$^{5}$ years.  Note that a less massive mid-B star would have a lifetime about 5$\times$ longer, changing our absolute lifetime estimates by that factor.  This variation gives a sense of the uncertainties involved in deriving absolute lifetimes.}, $\tau_{UCHII}$.  
The remaining link between the absolute and relative lifetimes is the fraction of DMRs associated with UCHII regions, particularly O stars.  \citet{woo89} look for only the brightest UCHII regions, dense regions containing massive stars, while the more recent and more sensitive studies of \citet[e.g.][]{and11} show HII regions over wider evolutionary stages, after much of the dense gas cocoon has been dispersed.  We suggest that future works investigate the use of newer catalogs from the CORNISH survey \citep[e.g.;][]{pur13}.  \citet{woo89} searched three regions in our $l$ = 30\deg\  field (all classified as ``starry" in our study) for UCHII regions and found them toward two.  Therefore the \citet{woo89} absolute lifetime of 3.6 $\times$ 10$^{5}$ years ($\tau_{UCHII}$) corresponds, very roughly, to 2/3 of `starry' DMRs they surveyed in our analyzed field. 

We make the assumption that approximately 50-100\% of our `starry' pixels are associated with UCHII regions ($f_{UCHII}$).  This assumption is based on the following lines of evidence: 1) 8 \micron~emission is often indicative of UV excitation \citep[show that nearly all GLIMPSE bubbles, 8 \micron~emission, are associated with UCHII regions]{ban10}, 2) ``starry" pixels show warmer dust temperatures, 3) and UCHII regions were found toward 2/3 regions surveyed by \citet{woo89} in our field, as shown above. 
The assumption that 50-100\% of `starry' pixels are associated with UCHII regions ($f_{UCHII}$) corresponds to total DMR lifetimes ($\tau_{total}$) of 2.4 or 1.2 Myr (for 50\% or 100\%, respectively) when we assume a starry fraction of 30\% ($f_{starry}$) as shown in the equation below.
\begin{equation}
\tau_{total} = \frac{\tau_{UCHII}}{f_{starry}~f_{UCHII}}
\end{equation}
The absolute lifetimes of the `Generous' and `Conservative' thresholds starless phase (using Equations 5 and 6) is then 0.8-1.7 Myr and the ``starry" phase would be 0.4-0.7 Myr.  
If we instead assume the KP10 column density threshold for high-mass star formation, $f_{UCHII}$ would not change, only the star-forming and starless fraction, in this case, 37\% and 63\%, respectively.  The absolute lifetimes inferred for these phases based on association with UCHII regions is a $\tau_{total}$ of 1.0-1.9 Myr, a $\tau_{starless}$ of 0.6-1.2 Myr and a $\tau_{starry}$ of 0.4-0.7 Myr.  See \autoref{table-comp} for a summary of the relative and absolute lifetimes for various methods.

\subsection{Free-fall times}
\label{sec:ff}

The absolute lifetimes derived in \S \ref{sec:maser} and \ref{sec:uchii} are compared with fiducial cloud free-fall times.  To calculate `fiducial' cloud free-fall times, we first calculate the median pixel column density for each of the categories (conservative, generous, KP10 near, and KP10 far).  These are shown in \autoref{table-ff}.  We then use Equations 11 and 12 from \citet{svo16} to calculate a fiducial free-fall time from the column densities.  The central volume density is calculated from the column density assuming a characteristic length of 1 pc and a spherically symmetric  Gaussian density distribution \citep[see][for details]{svo16}.  We stress that there are many uncertainties in this calculation, as we do not know the true volume densities, but these are simply meant to provide approximate free-fall times based on known cloud parameters, such as the median column density and typical size.  Moreover, these free-fall times are, of course, calculated at the `present day,' and we do not know what the `initial' cloud free-fall times were.  The fiducial free-fall times for each category are listed in \autoref{table-ff}.

For each category, we then convert the absolute lifetimes derived into number of free-fall times.  These are shown in the rightmost three columns of \autoref{table-ff}.  The number of free-fall times for the total lifetimes range from 1-6.5.  The starless phase ranges between 0.6-4.1 free-fall times and the starry phase ranges from 0.4-2.4 free-fall times.

\begin{table*}
\begin{center}
\begin{tabular}{ccccc}
\hline
Clump Identification & SF Identification & $f_{starless}$ &  Reference \\
\hline\hline
IRDC & 24 \micron & 0.65 & \citet{cha09} \\
IRDC & 8 \micron & 0.82 & \citet{cha09}\\
IRDC & 24 \micron & 0.33 & \citet{par09} \\
IRDC & 24 \micron & 0.32-0.80 & \citet{per09}\\
LABOCA & 8/24 \micron & 0.44 & \citet{mie12} \\
ATLASGAL & 22 \micron &  0.25  &  \citet{cse14} \\
ATLASGAL & GLIMPSE+MIPSGAL & 0.23 & \citet{tac12} \\
Hi-GAL & 24 \micron & 0.18 & \citet{wil12} \\
Hi-GAL & 8 \micron & 0.33 & \citet{wil12} \\
BGPS & mid-IR catalogs\footnote{\citet{rob08}, the Red MSX Catalog from \citet{urq08}, and the EGO catalog from \citet{cyg08}.} & 0.80 & \citet{dun11a} \\
BGPS & many tracers & 0.47 & \citet{svo16} \\
ATLASGAL & MIPSGAL YSOs & 0.69 & \citet{hey16} \\ 
N(H$_{2}$) from Hi-GAL & T$_{dust}$ and 8 \micron & 0.60-0.70 & This work \\
\hline
\end{tabular}
\end{center}
\caption{Previous estimates for the starless phase fraction, $f_{starless}$.}
\label{table-sf}
\end{table*}

\subsection{Comparison with Other Lifetime Estimates for Dense, Molecular Clumps}
\label{sec:comp}

Previous lifetime estimates are summarized in \autoref{table-sf}.  They are based primarily on mid-IR emission signatures at 8/24 \micron~toward IRDCs or dust-identified clumps and find starless fraction percentages between 23-82\%.  In previous studies, these starless fractions are often extrapolated to absolute starless lifetimes ranging from $\sim$ 10$^{3}$-10$^{6}$ years.  %
Additionally, \citet{tac12} find a lifetime for the starless phase for the most high-mass clumps of 6 $\times$ 10$^{4}$ years based on an extrapolated total number of starless clumps in the Milky Way and a Galactic SFR, and \citet{ger14} found an IRDC lifetime of 10$^{4}$ years based on chemical models.


Previous analyses determined lifetimes of high-mass star-forming regions by calculating relative fractions of `clumps' or `cores,' defined in various ways and with arbitrary sizes.  All the regions within each `clump' or `core' are lumped together and collectively denoted as starless or star-forming.  Since clumps identified in sub-mm surveys generally contain gas in different stages of star formation (actively star-forming and quiescent) we chose to use the pixel-by-pixel approach.  In this way, gas that is star-forming or quiescent is identified as such without being instead included in a different category due to its association with a clump.  Previously, a single 8 or 24 \micron~point source would classify an entire clump as star-forming, therefore, we expect that our pixel-by-pixel approach will identify more regions as starless and give a higher starless fraction.  
Our relative lifetime estimates are in reasonable agreement with previous work on the topic, and yield a somewhat higher starless fraction, as would be expected with the pixel-by-pixel method.

Of particular interest for comparison are the recent works of \citet{svo16} and \citet{hey16}.  \citet{svo16} perform a comprehensive analysis of over 4500 clumps from BGPS across the Galactic Plane, including their distances, physical properties, and star-forming indicators.  In this analysis they notice a possible trend in which the clump starless lifetimes decrease with clump mass.  Overall, about 47\% of their clumps can be qualified as `starless' clump candidates, and using a similar method for determining absolute lifetimes, they find lifetimes of  0.37 $\pm$ 0.08 Myr (M/10$^{3}$ \Msun)$^{-1}$ for clumps more massive than 10$^3$ \Msun.  \citet{hey16} similarly perform a comprehensive analysis of the latency of star formation in a large survey of about 3500 clumps identified by ATLASGAL.  They carefully identify MIPSGAL YSOs \citep{gut15} that overlap with these clumps, and accounting for clumps excluded due to saturation, find that about 31\% are actively star-forming.  They conclude that these dense, molecular clumps have either no star formation, or low-level star formation below their sensitivity threshold, for about 70\% of their lifetimes.   Our starless lifetime of about 60-70\% agrees remarkably well with both of these studies.  The \citet{svo16} analysis includes regions of lower column density than our selection and are also sensitive to the early signatures of lower-mass stars, so it would be expected that the starless fraction is a bit lower. 

The absolute lifetimes we derive are larger than most previous studies simply because of how they are anchored - most previous studies simply assume a star-forming lifetime ($\tau_{starry}$) of 2 $\times$ 10$^{5}$ years \citep[representative YSO accretion timescale,][]{zin07}.  If this star-forming lifetime ($\tau_{starry}$) is used along with our starless fractions of 0.6-0.7, we would derive starless lifetimes ($\tau_{starless}$, using Equations 5 and 6) of 0.3 - 0.5 Myr. 

Overall, there is quite a wide range in the estimates of the starless lifetimes for DMRs.  However, the relatively good agreement from the comprehensive studies of \citet{svo16} and \citet{hey16} on individual clumps and the present work using a variety of star-formation tracers and a pixel-by-pixel analysis over a large field may indicate that these values are converging.  Moreover, it is crucial to understand that different techniques will necessarily provide different values, as each is probing a different clump or DMR density and some star-formation tracers will be more sensitive to the signatures of lower-mass stars.  The matter is overall, quite complex, and assigning a single lifetime to regions of different masses and densities is a simplification \citep{svo16}.

\subsection{Comparison with Global Milky Way SFRs}
\label{sec:compsfr}
One simple sanity test for our lifetime estimates is to compare them with global SFRs.
We use our column density map of the ``starry" regions and convert it to a total mass of material in the star-forming phase.  For the range of column densities considered, assuming distances between 5-9 kpc, we find the total mass of material engaged in forming high-mass stars to be about 0.5 - 3 $\times$ 10$^5$ $\Msun$ in the 2\deg~$\times$ 2\deg~field centered at [$\ell$, b] = [30\deg, 0\deg].  

We assume a typical star formation efficiency of 30\% to derive the mass of stars we expect to form in the region over ``starry" lifetime of 0.1 - 0.7 Myr.  Multiplying the gas mass by the efficiency and dividing by this lifetime gives us a total SFR in our 2\deg~$\times$ 2\deg~region of 0.02 - 0.82 \Msun/year (given the ranges in lifetimes, threshholds and distances).  Extrapolating this to the entire Galaxy within the solar circle \citep[assuming a sensitivity of our measurements from 3 to 12 kpc and a typical scale height of 80 pc as in][]{rob10} gives a global Milky Way SFR ranging from 0.3 - 20 \Msun/year.  Typical estimates of Milky Way SFRs range from about 0.7 - 10 \Msun/year, and when accounting for different IMFs, converge to about 2 \Msun/year \citep[e.g.][]{rob10, cho11}.  

Since our observed field contains W43, often touted as a ``mini-starbust" \citep[e.g.][]{bal10,  lou14}, we expect our inferred global SFR to be higher than the true global SFR.  Due to the many uncertainties and assumptions, we find that the inferred global SFR has a large range and is between 0.1 - 10 $\times$ the fiducial value of 2 \Msun/year.  While the level of uncertainties and assumptions preclude any meaningful inference from this comparison, our numbers do pass the simple sanity check.  Additionally, the average of inferred global SFRs is higher than the fiducial value, as would be expecåted for this highly active, ``mini-starburst" region of the Galaxy.

\section{Conclusion}
\label{sec:conclusion}
We estimate the relative lifetimes of the starless and star-forming phases for all regions capable of forming high-mass stars in a 2\deg~$\times$ 2\deg~field centered at [$\ell$, b] = [30\deg, 0\deg].  We use column densities derived from Hi-GAL to determine which regions are capable of forming high-mass stars, and dust temperature and Spitzer 8 \micron~emission to determine if the region is starless (to a limit of about 10 \Msun) or star-forming.  Unlike previous analyses, we do not create any artificial `clump' boundaries, but instead use all the spatial information available and perform our analysis on a pixel-by-pixel basis.   We find that regions capable of forming high-mass stars spend about 60-70\% of their lives in a starless or embedded phase with star formation below our detection level and 30-40\% in an actively star-forming phase.

Absolute timescales for the two phases are anchored to the duration of methanol masers determined from \citet{van05} and the UCHII region phase from \citet{woo89}.  
We include a wide range of possible assumptions and methodologies, which gives
a range for starless lifetimes of 0.2 to 1.7 Myr (60-70\%) and a star-forming lifetime of 0.1 to 0.7 Myr (30-40\%) for high-mass star-forming regions identified in the dust continuum above column densities from 0.4 - 2.5 $\times$ 10$^{22}$ cm$^{-2}$.  These lifetimes correspond to about 0.6-4.1 free-fall times for the starless phase and 0.4-2.4 free-fall times for the star-forming phase, using fiducial cloud free-fall times.  In this work, we are only sensitive to tracing forming stars more massive than about 10 \Msun.  If lower-mass stars in the same regions form earlier, the starless timescale for those stars would be even faster than the 0.6-4.1 free-fall times reported here. 

We find that the relative lifetimes of about 60-70\% of time in the starless phase and 30-40\% in the star-forming phase are robust over a wide range of thresholds, but that the absolute lifetimes are rather uncertain.  These results demonstrate that a large fraction of high-column density gas is in a starless or embedded phase.  We outline a methodology for estimating relative and absolute lifetimes on a pixel-by-pixel basis. This pixel-by-pixel method could easily be implemented to derive lifetimes for dense, molecular regions throughout the Milky Way.

\acknowledgments
We thank the anonymous referee for many insightful and important comments that have greatly improved the manuscript.  We also thank P. Meyers, Y. Shirley, H. Beuther, J. Tackenberg, A. Ginsburg, and J. Tan for helpful conversations regarding this work.  
Data processing and map production of the Herschel data has been possible thanks to generous support from the Italian Space Agency via contract I/038/080/0. 
Data presented in this paper were also analyzed using The Herschel interactive processing environment (HIPE), a joint development by the Herschel Science Ground Segment Consortium, consisting of ESA, the NASA Herschel Science Center, and the HIFI, PACS, and SPIRE consortia.  
This material is based upon work supported by the National Science Foundation under Award No. 1602583 and by NASA through an award issued by JPL/Caltech via NASA Grant \#1350780.

\bibliography{references1}{}

\end{document}